\documentclass[aps,prb,twocolumn,showpacs,amsmath,amssymb,superscriptaddress,footinbib,bibnotes]{revtex4-1}

\usepackage{bm}
\usepackage{graphicx,color}
\usepackage[T1]{fontenc}
\usepackage{hyperref}

\def\cA{{\mathcal A}}
\def\bcA{\boldsymbol{\mathcal A}}

\def\br{{\bf r}}
\def\bj{{\bf j}}
\def\r]{\right]}
\def\l[{\left[}

\def\ap{\alpha_{+}}
\def\am{\alpha_{-}}

\newcommand{\augsburg}{Universit\"at Augsburg, Institut f\"ur Physik, 86135 Augsburg, Germany}
\newcommand{\saclay}{Service de Physique de l'Etat Condens\'{e}, 
                     CNRS URA 2464, CEA Saclay, 
                     91191 Gif-sur-Yvette, France}

\begin{document}

\title{Driving spin and charge in quantum wells by surface acoustic waves}

\author{Johannes Wanner}
\email{johannes.wanner@physik.uni-augsburg.de}
\affiliation{\augsburg}

\author{Cosimo Gorini}
\email{cosimo.gorini@cea.fr}
\affiliation{\saclay}

\author{Peter Schwab}
\affiliation{\augsburg}

\author{Ulrich Eckern}
\email{ulrich.eckern@physik.uni-augsburg.de}
\affiliation{\augsburg}



\begin{abstract}
Recent experiments have shown the potential of surface acoustic
waves as a mean for transporting charge and spin in quantum wells.
In particular, they have proven highly effective for the coherent
transport of spin-polarized wave packets, suggesting their potential in
spintronics applications.  Motivated by these experimental observations,
we have theoretically studied the spin and charge dynamics in a quantum
well under surface acoustic waves.  We show that the dynamics acquires a simple and transparent
form in a reference frame co-moving with the surface acoustic wave. Our results, e.g.,
the calculated spin relaxation and precession lengths, are in excellent agreement with 
recent experimental observations.
\end{abstract}

\maketitle

\section{Introduction}
\label{sec_intro}
Coherent spin transport across a device is a central goal
of spintronics.\cite{Stotz2005,Awschalom2007} In this context the enhancement of the spin lifetime
is a critical issue, and recent experiments have demonstrated the effectiveness 
of using Surface Acoustic Waves (SAWs) for this purpose. 
\cite{Sogawa2001,Stotz2005,Couto2007,Couto2008,Sanada2011}
In such experiments the spin density in semiconducting quantum wells is optically generated by laser beams
and transported by a SAW over distances of several tens of micrometers.
The current understanding\cite{Stotz2005,Couto2008} is that these long distances are possible 
due to the suppression of both Bir-Aronov-Pikus\cite{Bir1975} 
and Dyakonov-Perel'\cite{Dyakonov1972} spin-relaxation mechanisms: 
the piezoelectric SAW potential spatially separates electrons and holes,
thus inhibiting Bir-Aronov-Pikus relaxation, and at the same time
confines them to narrow (moving) wires/dots, which causes motional narrowing and thus a suppression
of Dyakonov-Perel' relaxation.  However, motional narrowing in a 2-Dimensional Electron Gas (2DEG)
ceases to be relevant for strong \textit{static} confinements, when spin-dependent scattering at the boundaries
takes over, as recently observed\cite{Holleitner2006} and theoretically explained.\cite{Schwab2006}
In this work we address the question of \textit{dynamic} confinement.
In particular, we will investigate how intrinsic (Dyakonov-Perel')
spin relaxation mechanisms affect the spin dynamics of pockets of photoexcited electrons
driven by SAWs.  

We will also briefly comment on the role of extrinsic (Elliot-Yafet) spin relaxation.\cite{Raimondi2009}
Spin relaxation due to the hyperfine interaction between the carriers and the background nuclei
may be an important issue in strongly confined, static geometries \cite{Merkulov2002,Braun2005}, 
but was recently shown\cite{Echeverria2013} to be irrelevant for a pocket of mobile electrons
carried by a SAW, and hence will not be considered here.

We will start in Secs.~\ref{sec_model} and \ref{sec_diff} by defining the model
and introducing the diffusive limit, respectively.  In Sec.~\ref{sec_charge} charge dynamics will be discussed,
and in Sec.~\ref{sec_spin} the central issue of spin dynamics. For the sake of clarity, the latter will be studied by specializing to a specific geometry, and by retaining only the dominant spin-orbit interactions.
In Sec.~\ref{sec_superplus} we will comment on different geometries and additional spin-orbit terms. A short summary is given in Sec.~\ref{sec_conclusions}.


\section{The model}
\label{sec_model}

We consider an electron gas in the $x$-$y$--plane described by the Hamiltonian
\begin{equation}
H = \frac{p^2}{2m}+H_{\rm so}+V(\br).
\end{equation}
Here $m$ is the effective mass, $H_{\rm so}$ describes intrinsic spin-orbit coupling,
and $V(\br)$ is the random impurity potential. For the latter, we assume the standard ``white noise''
disorder, i.e., we assume that the average of the potential is zero, and its correlations are given by 
\begin{eqnarray}
\langle V(\br)V(\br')\rangle = (2\pi N_0\tau)^{-1}\delta(\br-\br').
\end{eqnarray}
Here $N_0 = m/(2\pi)^2$ is the density of states at the Fermi energy per spin, $\tau$ is the elastic
momentum scattering time, and we have chosen $\hbar=1$.

For $H_{\rm so}$ we consider general linear-in-momentum couplings,
which arise in 2DEGs because of broken structural (Rashba\cite{Bychkov1984}) or bulk 
(Dresselhaus\cite{Dresselhaus1955}) inversion symmetry, or of strain;\cite{Bernevig2005}
linear couplings are dominant with respect to cubic ones in a wide range of parameters.\cite{Studer2010, Walser2012}
Linear-in-momentum couplings can be written in terms of a non-Abelian vector potential 
$\bcA$,\cite{Mathur1992,Frohlich1993,Tokatly2008,Gorini2010}
which for spin $1/2$ carriers becomes a $SU(2)$ field with three components 
in the Pauli matrices basis ($a=x,y,z$), and two components in real space ($i=x,y$):
\begin{equation}
\label{su2}
H_{\rm so} = p_i\cA_i^a\sigma^a/2m.
\end{equation}
Unless otherwise specified, upper (lower) indices will refer to spin (real space) components
throughout.  

Our treatment is based on the general approach described in Refs.~\onlinecite{Gorini2010,Raimondi2012}. 
However, for definiteness we will start by considering quantum wells grown 
in the $\hat{\bf z}\parallel [001]$ direction. 
With the in-plane base vectors $\hat{\bf x}\mid\mid [100]$ and $\hat{\bf y}\mid\mid [010]$ 
the linear Rashba and Dresselhaus spin-orbit Hamiltonians read
\begin{eqnarray}
\label{eq_hR}
H^R_{\rm so} &=& \alpha (p_y\sigma^x-p_x\sigma^y),
\\
\label{eq_hD}
H^D_{\rm so} &=& \beta (p_y\sigma^y-p_x\sigma^x),
\end{eqnarray}
with $\alpha, \beta$ the respective coupling constants. 
These spin-orbit terms can be rewritten according to \eqref{su2} with the following $SU(2)$ potentials:
\begin{eqnarray}
\label{eq_AR}
&(\cA_R)^x_y = - (\cA_R)^y_x = 2m\alpha ,&
\\
\label{eq_AD}
&(\cA_D)^y_y = - (\cA_D)^x_x = 2m\beta&,
\end{eqnarray}
all other components being zero. 
 
The spin-orbit interaction depends on the electron direction of motion;
thus, in order to examine the effect of a SAW, we will consider the latter 
to be propagating either in the $[110]$ or in the $[\bar{1}10]$ direction. 
In both cases the driving field can be written as
\begin{equation}
\label{drivingSAW}
{\bf E}(\br)=E\,\hat{\mathbf{k}}\cos\left(\mathbf{kr} -\omega t\right), 
\end{equation}
where $\omega=v \vert\mathbf{k}\vert$; $v$ is the sound velocity in the medium, and $ \hat{\mathbf k} $ 
is the unit vector pointing in the SAW propagation direction.

The SAW is generated in a piezoelectric material by applying a time-modulated voltage 
to interdigital transducers in contact with it, and the in-plane field \eqref{drivingSAW} 
is accompanied by a component in the $z$ direction and by strain.\cite{Mamishev2004,Morgan2007}
The latter are both sources of additional non-homogeneous and time dependent spin-orbit terms 
in the Hamiltonian.\cite{Sanada2011}  We will at first neglect these complications, and start by taking into account 
only the driving SAW field \eqref{drivingSAW}.


\section{Diffusive limit}
\label{sec_diff}

Within the $SU(2)$ ``color'' approach,\cite{Tokatly2008,Gorini2010,Raimondi2012} 
the charge and spin dynamics can be described by the $SU(2)$-covariant continuity equation
\begin{equation}
\label{eq_cont}
 \frac{\partial \rho}{\partial t}+\tilde{\nabla}\cdot\bj =0, 
\end{equation}
with the density and current given by
\begin{equation}
\rho=\rho^{0}+s^{a}\sigma^{a},\quad \bj=\bj^{0}+\bj^{a}\sigma^{a}.
\end{equation}
Here $ \rho^0 $ and $s^a $ are, respectively, the charge and spin ($a$-th component) density.
The covariant derivative 
\begin{equation}
\tilde{\nabla} =\nabla + \mathrm{i}\left[\bcA,...\right], 
\end{equation}
where 
\begin{equation}
\bcA= \left(\bcA^x \sigma^x +\bcA^y \sigma^y+\bcA^z \sigma^z\right)/2
\end{equation}
is defined according to Ref. \onlinecite{Gorini2010},
consists of two terms, the spatial derivative $\nabla$ and the commutator with the vector potential 
describing spin precession around the spin-orbit field. In this work, we consider the diffusive regime, i.e., we assume that the mean free path, $ l= v_F \tau $, is much smaller than the wavelength of the SAW, $ 2\pi/k $.  
In this limit, the electric field $ \textbf{E} $  of the SAW enters the charge-spin current as follows:\cite{Gorini2010}
\begin{equation}
\label{eq_current}
\bj=-D \tilde{\nabla}\rho +\mu \textbf{E}\rho,
\end{equation}
where $ D $ is the diffusion constant, and $ \mu $ the mobility.  
This simple structure is due to the fact that we are dealing with linear-in-momentum spin-orbit interactions.  
Substituting \eqref{eq_current} into the continuity equation \eqref{eq_cont} leads 
to a drift-diffusion equation for the charge density $\rho^0$, and to Bloch-type equations for the spin densities $s^a$.


\section{Charge dynamics}
\label{sec_charge}

As discussed above, the drift-diffusion equation 
for the charge carriers in the diffusive limit has the well known form:
\begin{equation}
 \frac{\partial \rho^{0}}{\partial t}+\mu\nabla\cdot(\textbf{E}\rho^{0}) -D\nabla^{2}\rho^{0}=0.
\label{dyn}
\end{equation}
In the following we assume the $ x $ axis to be parallel 
to the SAW propagation direction.  Since there is no drift of the carriers in the direction perpendicular 
to the SAW ($ y $ axis), the solution of the drift-diffusion equation factorizes,
$ \rho_0(\textbf{r},t)=a_0 X(x,t) Y(y,t) $.
Here $a_0$ is a constant fixed by the initial conditions, irrelevant for the dynamics
and thus neglected in the following unless otherwise specified.
The motion in the $ y $ direction is governed by the solution of the diffusion equation,
\begin{equation}
Y(y,t)=\frac{1}{\sqrt{4\pi D t}}\int_{-\infty}^{\infty}\,\mathrm{d}y'\exp\left[-\frac{(y-y')^2}{4 D t}\right]
Y(y',0).
\label{ydens}
\end{equation}
For the dynamics in $x$ direction one has to discriminate between two cases, 
depending on the SAW velocity $ v $ being larger or smaller than the carrier velocity $ \mu E $.  
In the first case, $ v>\mu E $, the carriers are too slow to follow the SAW, but 
move from one minimum to the next.
Considering in addition not too small $E$ such that $Dk\ll\mu E$, cf.\ Eq.~(\ref{dyn}),
diffusion can be neglected and the dynamics is governed by the drift. In typical non-degenerate
semiconductors the Einstein relation $D = \mu k_B T/e$ can be employed to estimate the diffusion
constant,\cite{Cameron1996, Wang2013, Garcia-Cristobal2004} implying that the condition
$Dk\ll\mu E$ becomes independent of the mobility, namely reduces to $k_B T \cdot k \ll eE$, or
$k_B T\ll eE/k$. This requirement is easily met at low temperatures, $T \sim 20$ K or
lower.\cite{Garcia-Cristobal2004} Though diffusion acquires importance with increasing temperature,
the experimental data of Ref.~\cite{Couto2008} (see Fig.\ 4(b) therein), where $k\approx 1.12 \times
10^4 {\rm cm}^{-1}$ and $eE \approx 3.4 \times 10^3$ eV/cm, show that drift can be dominant even at
room temperature. In this case, the differential equation (\ref{dyn}) simplifies, 
and $ X(x,t) $ is found to be given by
\begin{equation}
X(x,t)=\frac{v-\mu E\cos\left[k \,\xi(x,t)\right]}{v-\mu E \cos\left(k x-\omega t\right)}X\left(\xi(x,t),0\right),\label{dde}
\end{equation}
with
\begin{widetext}
\begin{equation}
\xi(x,t)=\frac{2}{k} \arctan 
\left\lbrace\sqrt{\frac{v-\mu E }{v+\mu E }}
\tan\left[\arctan\left(\sqrt{\frac{v+\mu E }{v-\mu E}}\tan\left(\frac{k x-\omega t}{2}\right)\right)
+\frac{\sqrt{v^{2}-(\mu E )^{2}}}{2v}\,\omega t \right] 
\right\rbrace.
\end{equation}
\end{widetext}
Note that $ \xi(x,t=0)=x. $

\begin{figure}
\includegraphics[width=0.45\textwidth]{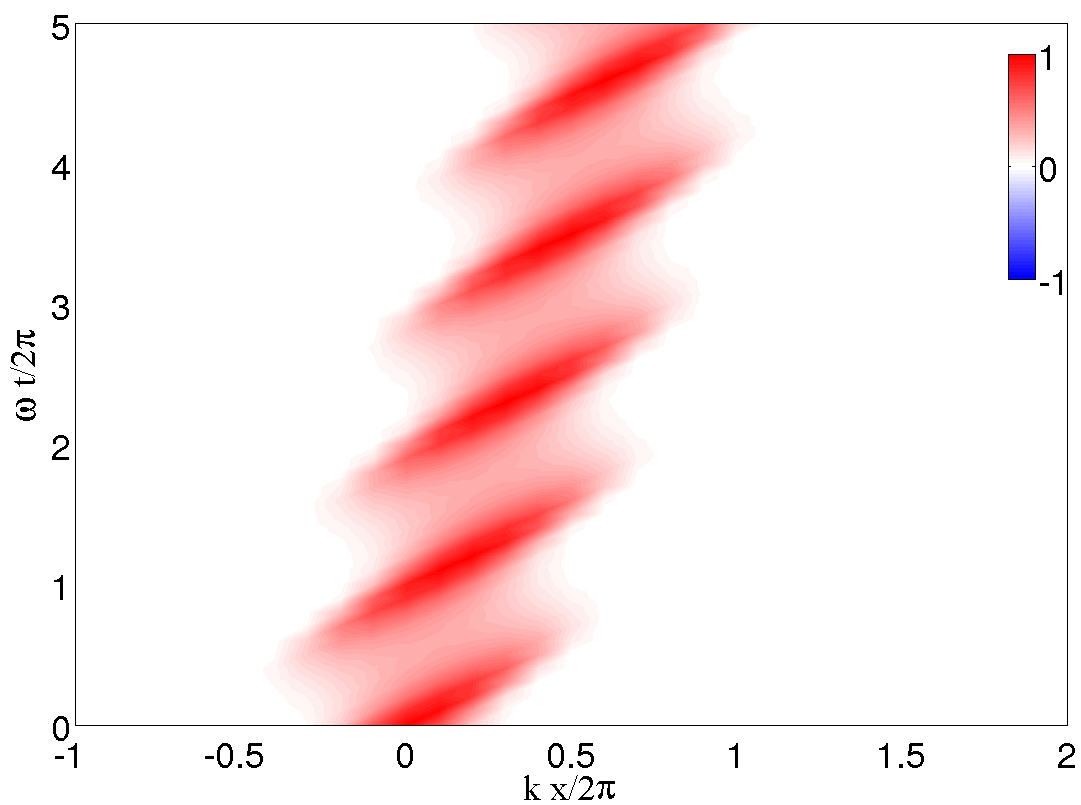}
\caption{Motion of the charge carriers $ X(x,t) $ in $ x $ direction with $ \mu E/v=0.5 $.}
\label{fig_charge}
\end{figure}

Care is needed because of the periodicity of $ \tan\left[\left(k x-\omega t\right)/2\right] $, since for an arbitrary initial condition one has to choose the right branch in order to obtain the solution with the correct initial distribution. One can circumvent this difficulty by choosing an initial condition with all carriers within one period. In Fig. \ref{fig_charge} we therefore assumed a Gaussian initial distribution with a standard deviation much smaller than the SAW wavelength. Although the carriers are not fast enough to follow the SAW, they flow from one minimum to the next, with 
the average velocity
\begin{equation}
\overline{v}=v-\sqrt{v^{2}-(\mu E)^{2}}, \; \mu E < v.
\end{equation}

The situation is quite different for $ \mu E > v $, when the charge carriers are fast enough to follow the SAW, i.e. they are ``surfing''. This means that they are subjected to a stationary potential in a reference frame moving with the SAW, and at the point $ x_0 =\arccos\left(v/\mu E\right)/k $ in this frame they move with its velocity. \footnote{Note that $ k x_0 $ lies in the range $ 0\ldots\pi $. This is also apparent from Eq. \ref{eq_carrier}. }  
Since the potential is periodic, there is such a point in every period.  
Independent of the initial distribution $ X(x,0) $, the carriers flow towards 
the point $ x_0 $ corresponding to their period, until they reach a stationary distribution.  
Thus, for $ \mu E > v $ the solution (\ref{dde}) converges to $ X(x,t) \sim\delta (k (x-x_0)-\omega t) $, 
and the carrier density $ X(x,t) $ is concentrated in an infinitely small wire parallel 
to the wave front. This implies that the diffusion term cannot be neglected anymore.
Since the charge density distribution becomes stationary, the charge current vanishes in the moving frame, 
leading to
\begin{equation}
X(x,t)=\exp\left[\frac{\mu E \sin(kx-\omega t)-v (k x-\omega t) }{D k}\right],
\label{eq_carrier}
\end{equation}
which is sharply peaked at $ kx -\omega t=kx_0 $. Hence for $ \vert kx - \omega t -k x_0 \vert \ll 1 $, $ X(x,t) $ 
can be approximated by a Gaussian distribution, 
\begin{equation}
X(x,t)\approx\frac {e^{-\frac {(kx-\omega t-k x_0 )^2} {2\sigma^2}}} {\sqrt {2\pi}\sigma},
\end{equation}
with standard deviation $ \sigma^2= D k /\sqrt{(\mu E)^{2}-v^{2}} $.  Note that the exact solution 
(\ref{eq_carrier}) of the continuity equation (\ref{dyn}) does not depend on the sign of $ \mu $. 
In other words, this solution describes the dynamics of electrons as well as that of holes,
provided both are in the surfing regime, i.e., $\mu_{\rm e}E$, $\mu_{\rm h}E>v$.  
In this case the spatial separation of the two pockets of carriers is
\begin{equation}
\Delta x_0=\frac{\arccos\left(v/\mu_{\mathrm{h}} E\right)-\arccos\left(v/\mu_{\mathrm{e}} E\right)}{k}.\label{distance}
\end{equation}


\section{Spin dynamics}
\label{sec_spin}

In this section, we examine 
the influence of a SAW on the spin density. 
The spin-orbit Hamiltonians can be written as
\begin{eqnarray}
\label{newcoord1}
H^R_{\rm so}+H^D_{\rm so}=&-(\alpha+\beta)\frac{p_x-p_y}{\sqrt{2}}\,\frac{\sigma^x+\sigma^y}{\sqrt{2}}\\
&+(\alpha-\beta)\frac{p_x+p_y}{\sqrt{2}}\,\frac{\sigma^x-\sigma^y}{\sqrt{2}}\nonumber\\
\label{newcoord2}
=:&\ap p_{x'}\sigma^{y'}+\am p_{y'}\sigma^{x'}
\end{eqnarray}
where the primed coordinates correspond to the two directions, $ [110] $ and $ [\bar{1}10] $.  
In the following, we will perform all our calculations in this rotated reference frame 
(both real space and spin components rotated by $ \pi/4 $ around the $ z $ axis) 
with $ \hat{\mathbf{x}}\parallel [110] $ and $ \hat{\mathbf{y}}\parallel [\bar{1}10] $, but drop the prime (except for the closing of Sec.~\ref{sec_superplus} where we will revert back to non-rotated coordinates). For the vector potential $ \bcA $ one finds
\begin{eqnarray}
&(\cA)_x^y=-2m(\alpha+\beta):=\,2m\ap,&\\
&(\cA)_y^x=2m(\alpha-\beta):=\,2m\am,&\\
&(\cA)_x^x=(\cA)_y^y=0.& 
\label{vecpot}
\end{eqnarray}
The Bloch equations describing the dynamics of the spin density read
\begin{eqnarray}
\partial_{t} s^a+\mu\nabla \cdot\textbf{E}s^a - D \nabla^{2}s^a & = & -2 D\,
\epsilon_{abc}\,\boldsymbol{\mathcal{A}}^b \cdot\nabla s^c - \Gamma^{ab} s^b+\nonumber\\
& & + \epsilon_{abc}\,\mu\textbf{E}\cdot\bcA^b s^c .
\label{eq_bloch}
\end{eqnarray}
These set of equations are obtained by taking the spin $a$-component 
of the continuity equation (\ref{eq_cont}), after expressing the current in the diffusive regime
according to \eqref{eq_current}. Without a SAW and for a homogeneous spin distribution, 
one can immediately determine the spin lifetimes from the eigenvalues 
of the inverse spin relaxation matrix $ {\hat\Gamma}^{-1} $. In fact
$ \hat\Gamma $ in (\ref{eq_bloch}) is diagonal, and its eigenvalues are
\begin{eqnarray}
\Gamma_x &=&\, 4Dm^2 \ap^2 \,, \label{Gamma1} \\ 
\Gamma_y &=&\, 4Dm^2 \am^2 \,, \label{Gamma2} \\
\Gamma_z &=&\, 4Dm^2 \left(\ap^2+\am^2\right)\,. \label{Gamma3}
\end{eqnarray}
From \eqref{Gamma2} one sees that for $y$-polarized spins there is no relaxation 
if $ \alpha=\beta $.  Although this limit can be realized in experiments,\cite{Koralek2009,Kohda2012} 
we here consider the more general case $ \alpha\neq \beta $.

\subsection{Homogeneous initial conditions}
\label{subsec_homo}

The spin dynamics depends strongly on the initial conditions.  
In this subsection we consider an experimental setup where a short laser pulse homogeneously polarizes 
the complete surface.  In the surfing regime electrons and holes are strongly localized 
and effectively spatially separated, see Eq.~\eqref{distance},
and are transported---along with their spins---across the sample.  
The description of the spin dynamics is considerably simplified by switching to a reference frame 
co-moving with the SAW.  A change to such a reference frame leads to an additional term 
in the continuity equation which acts like an internal magnetic field, 
\begin{equation}
\partial_{t}\rho+\tilde{\nabla}\textbf{j}+\mathrm{i}[\textbf{v}\boldsymbol{\mathcal{A}},\rho]=0.
\end{equation}
where $ \textbf{v} = v \hat{\mathbf{k}}$.
A further simplification can be achieved by applying the following $SU(2)$ gauge transformation:
\begin{equation}
\boldsymbol{\mathcal{A}}\rightarrow U^{\dagger}\boldsymbol{\mathcal{A}}U+iU^{\dagger}\nabla U, \;
U=\exp\left(\mathrm{i}x\mathcal{A}_{x} \right).
\end{equation}
In this gauge, the covariant derivative $ \tilde{\partial}_x \rightarrow \partial_x $ is diagonal in
spin space but leads to a $ x $-dependent vector potential $ \mathcal{A}_{y}(x) $. However, since
the charge carriers are Gaussian-distributed at the origin in the co-moving system with
$ \sigma \ll 1/2m \ap $, one can neglect the $ x $-dependence of the vector potential,
hence $ \mathcal{A}_{y}(x)\approx\mathcal{A}_{y}(0) $.

The time-dependence of the spin density in the presence of the SAW is governed by an effective 
relaxation matrix $\hat\gamma$, whose (complex) eigenvalues are given by 
\begin{eqnarray}
\gamma_{x,z}& = & 2Dm^2 \am^2\pm2\,\mathrm{i}\sqrt{v^2 m^2 \ap^2-D^2 m^4 \am^4}\label{eig110-1},\\ 
\gamma_y & = & \, 4Dm^2 \am^2.
\label{eig110-2}
\end{eqnarray}
Since all carriers move with the same velocity $ v $, the real part of these eigenvalues 
is related to the spin decay length, 
\begin{equation}
L_s=\frac{v}{\Re (\gamma)},\label{decaylength}
\end{equation}
whereas the imaginary part determines the spatial precession length,
\begin{equation}
\lambda=\frac{v}{\Im(\gamma)}.
\label{precessionlength}
\end{equation}
For a SAW moving in the $ y $ direction we proceed in the same way.  
The carriers are then concentrated in a small wire parallel to the $ x $ axis.  
In this case one finds
\begin{eqnarray}
\gamma_x & = & \, 4Dm^2 \ap^2,\\
\gamma_{y,z} & = & 2Dm^2 \ap^2\pm2\,\mathrm{i}\sqrt{v^2 m^2 \am^2-D^2 m^4\ap^4}
\end{eqnarray}
which is obtained from Eqs.~\eqref{eig110-1} and \eqref{eig110-2} by interchanging $ x $ and $ y $ 
as well as $+$ and $-$.  

Comparing the real parts with Eqs.~\eqref{Gamma1} and \eqref{Gamma2}, 
one finds a maximal enhancement of the spin lifetime by a factor of $ 2 (\ap/\am)^2 $ for the $ x $ direction, 
and $ 2 (\am/\ap)^2 $ for the $ y $ direction (``motional narrowing''). Note that the real parts of $ \gamma_{x/z} $ and $ \gamma_{y/z} $ 
are by a factor of two smaller than their perpendicular counterparts, $ \gamma_y $ and  $ \gamma_x $, 
respectively.  These perpendicular counterparts, describing the relaxation of spins parallel 
to the SAW wave front, are not affected by the SAW in the simple case of a homogeneous spin density.

\begin{figure}[htbc]
\begin{center}
\includegraphics[width=0.45\textwidth]{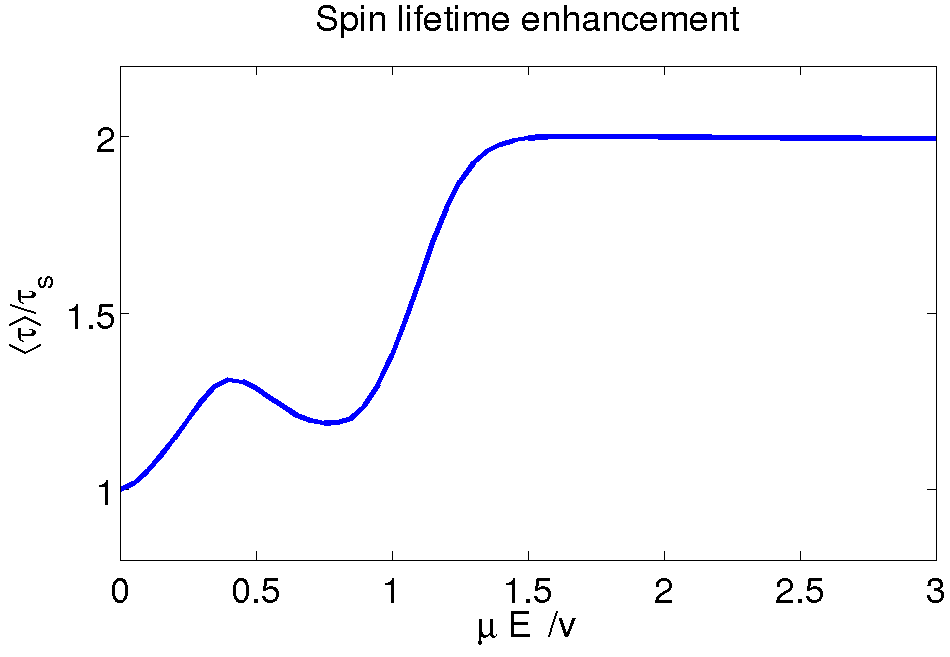}
\caption{Numerical results for the increase of the spin lifetime $ \langle\tau\rangle $ due to a SAW.  
For the calculation we assumed $ \ap=\am $.  
The spin lifetime is normalized by $ \tau_s=\Gamma^{-1}_x $, cf.~Eq.~\eqref{Gamma1}.}
\label{figspin}
\end{center}
\end{figure}

Specifically we numerically calculated the $x$-spin density for a SAW 
traveling in the $ x $ direction and for different $E$ values.  
For simplicity, we set $ \ap=\am $, which in the surfing regime implies a spin lifetime increase 
by a factor of two. Not being interested in the spatial variation of the spin density, 
we consider the spin polarization $ P_s=\vert\textbf{P}_s \vert $, by integrating 
the spin density over the whole surface.  From the Bloch equations \eqref{eq_bloch} 
one sees that, without a SAW, the spin polarization decays exponentially with the spin scattering rate
\eqref{Gamma1}. Hence we define the average spin lifetime by
\begin{equation}
\langle\tau\rangle=\frac{\int_{0}^{\infty}tP_s\,\mathrm{d}t }{\int_{0}^{\infty}P_s\,\mathrm{d}t }.
\end{equation}
For the numerical analysis, we started at $ t=0 $ with a Gaussian distribution in $ x $ direction, 
with a standard deviation $ \sigma \gg 1/2m\ap $, polarized in $ x $ direction.  
The spin lifetime as a function of the ratio $ \mu E/v $ is shown in Fig.~\ref{figspin},
where the expectation value $ \langle\tau\rangle $ is normalized to the corresponding spin lifetime 
$ \tau_s $ without SAW, cf.~Eq.~\eqref{Gamma1}.
In the regime $ \mu E/v <1 $, when the carriers are not surfing, the spin lifetime depends strongly 
on the form of the initial spin distribution; in particular, for our choice its
$E$-dependence is non-monotonic. 
As one approaches the surfing regime $ \mu E>v $ the spin lifetime converges 
to the expected value $ 2 \tau_s $.

\subsection{Inhomogeneous initial conditions}
\label{subsec_inhomo}

So far we have discussed the spin dynamics of an initially homogeneous spin distribution, for which
case there is no spin current parallel to the SAW wave front. However this assumption is not justified
in experiments where the initial spin distribution is created by, say, a focused laser beam. Again,
without loss of generality, we consider a SAW moving in $x$  direction.

While for the homogeneous case, the spins were precessing 
only around the axis parallel to the SAW wave front, now there will be diffusion along the wave front,
and hence they will also rotate around the SAW propagation direction. As a consequence the spins
along the narrow moving wire will not have the same orientation. In order to deal with this additional
precession we employ the following ansatz for the spin density:
\begin{equation}
s^a=\rho^0(r,\varphi,t)\,\eta^{a}(\varphi,t),
\label{ansatz}
\end{equation}
where $ r=2m\sqrt{\ap^2 x^2+\am^2 y^2} $ denotes the renormalized (dimensionless) radius, 
and $ \varphi=\arctan[\am y/(\ap x)] $. The carrier density in the surfing regime, $ \rho^0(r,\varphi,t) $,
was already determined in Sec.~\ref{sec_charge}, with $ X(x,t) $ given in \eqref{eq_carrier};
according to Eq.~\eqref{ydens} the carrier density along the $ y $ axis 
for an initial Gaussian distribution with standard deviation $ y_0 $ reads 
\begin{equation}
Y(y,t)=\frac{1}{\sqrt{2\pi(2 D t+y_0^2)}}\exp\left[-\frac{y^2}{2 (2 D t+y_0^2)}\right].
\end{equation}
Instead of switching to the SAW co-moving reference frame 
as in the homogeneous case, we stay in the laboratory frame but perform again
a gauge transformation,
\begin{equation}
\boldsymbol{\mathcal{A}}\rightarrow U^{\dagger}\boldsymbol{\mathcal{A}}U+iU^{\dagger}\nabla U,
\; U=\exp\left[\mathrm{i}(x-x_0-vt)\mathcal{A}_{x} \right],
\end{equation}
since as above all relevant spin dynamics takes place in a small wire parallel 
to the SAW wave front. With the ansatz \eqref{ansatz}, and by neglecting terms $ \mathcal{O}(r^{-1}) $, 
the continuity equation \eqref{eq_cont} reads
\begin{eqnarray}
\partial_{t}\eta-\mathrm{i}\,\frac{v}{\cos\varphi}\left[\mathcal{A}_x(\varphi),\eta\right]+
D\left[\mathcal{A}_y,\left[\mathcal{A}_y,\eta\right]\right]=0,
\label{2ddyn}
\end{eqnarray}
where
$ \mathcal{A}_x(\varphi)=\exp\left(-\mathrm{i}\frac{\varphi}{2}\sigma^z\right)\mathcal{A}_x
\exp\left(\mathrm{i}\frac{\varphi}{2}\sigma^z \right) $
is the vector potential rotated around the $ z $ axis. 
The second term in Eq.~\eqref{2ddyn} leads to spin precession around the $ \varphi $-dependent 
vector potential $ \mathcal{A}_x(\varphi) $, whereas the third term is responsible 
for the relaxation of the spin components perpendicular to the $ x $ axis. 
\begin{figure}[htbc]
\begin{center}
\includegraphics[width=0.45\textwidth]{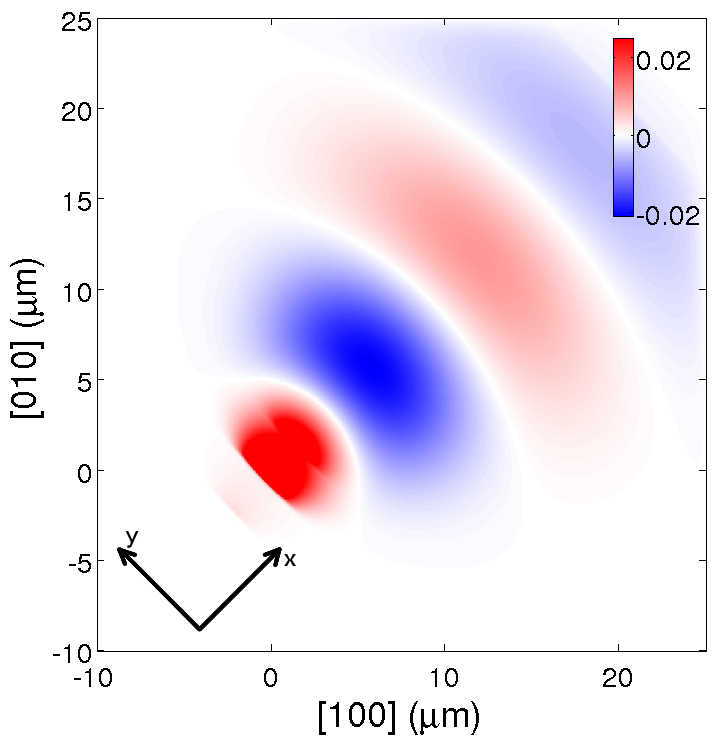}
\caption{Time-integrated spin density, $ \overline{s}^z $, for a SAW moving in $ [110] $ direction}\label{2D110}
\label{fig_110}
\end{center}
\end{figure}
\begin{figure}[htbc]
\begin{center}
\includegraphics[width=0.45\textwidth]{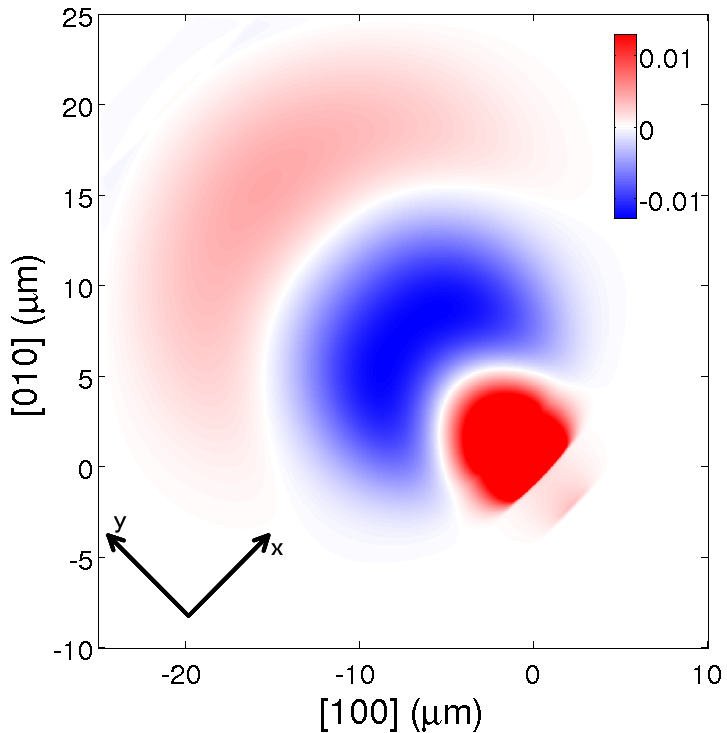}
\caption{Time-integrated spin density, $ \overline{s}^z $, for a SAW moving in $ [\bar{1}10] $ direction}\label{2D-110}
\label{fig_bar110}
\end{center}
\end{figure}
The Bloch equations now read
\begin{equation}
\partial_{t}\eta^a=- \gamma(\varphi)^{ab}\eta^b,
\end{equation} 
with the $\varphi$-dependent effective relaxation matrix
\begin{eqnarray}
&\\ \nonumber
\hat\gamma(\varphi)=&\left(\begin{array}{ccc}
0 & 0 & 2m v \ap  \\ 
0 & 4 Dm^2 \am^2   & 2m v  \ap \tan\varphi \\ 
- 2m v \ap  & -2m v  \ap \tan\varphi & 4 Dm^2 \am^2
\end{array} \right).
\end{eqnarray}
Assuming that the temporal resolution is not high enough to measure 
the time dependence of the spin density directly (see, e.g., Ref.~\onlinecite{Sanada2011}), we characterize
the additional rotation of the spins due to the diffusion parallel to the SAW wave front by the
time-integrated spin density: note that all spins are confined within a narrow wire, and the spin
density vanishes everywhere but for $ x-x_0 \approx vt $. For the time-integrated
spin density we therefore obtain
\begin{equation}
\overline{s}^a=\int_{0}^{\infty} s^a\,\mathrm{d}t\simeq a_0 Y\left(y,(x-x_0)/v\right)\eta^a\left(r,\varphi\right).
\end{equation}
The results presented in Figs.~\ref{2D110} and \ref{2D-110} were obtained by calculating
numerically the time-dependence of the spin density $ s^z $,
assuming at $t=0$ a Gaussian distribution with standard deviation of 1 $\mathrm{\mu m}$.  
Specifically, Figs.~\ref{2D110} and \ref{2D-110} show the time-integrated spin density for a SAW moving 
in $x$ and $y$ direction, respectively. We have chosen parameters comparable 
to the experimental ones\cite{Sanada2011} (we restore temporarily $\hbar$), namely
$ 2 m \alpha/\hbar^2 = 0.02 \,\mu \mathrm{m}^{-1} $, $ 2 m \beta/\hbar^2 = 0.17\, 
\mu  \mathrm{m}^{-1} $, $ D=30\, \mathrm{cm}^2/\mathrm{s} $, and $ v = 2.9 \times 10^5\,\mathrm{cm}/\mathrm{s} $.
The elliptical shape of the time-integrated spin density,
which is a consequence of the $ \varphi $-dependence of $ \mathcal{A}_x(\varphi) $,
is clearly visible in both figures, in remarkable agreement with the observed behavior.\cite{Sanada2011}

\begin{figure}[htbc]
\begin{center}
\includegraphics[width=0.45\textwidth]{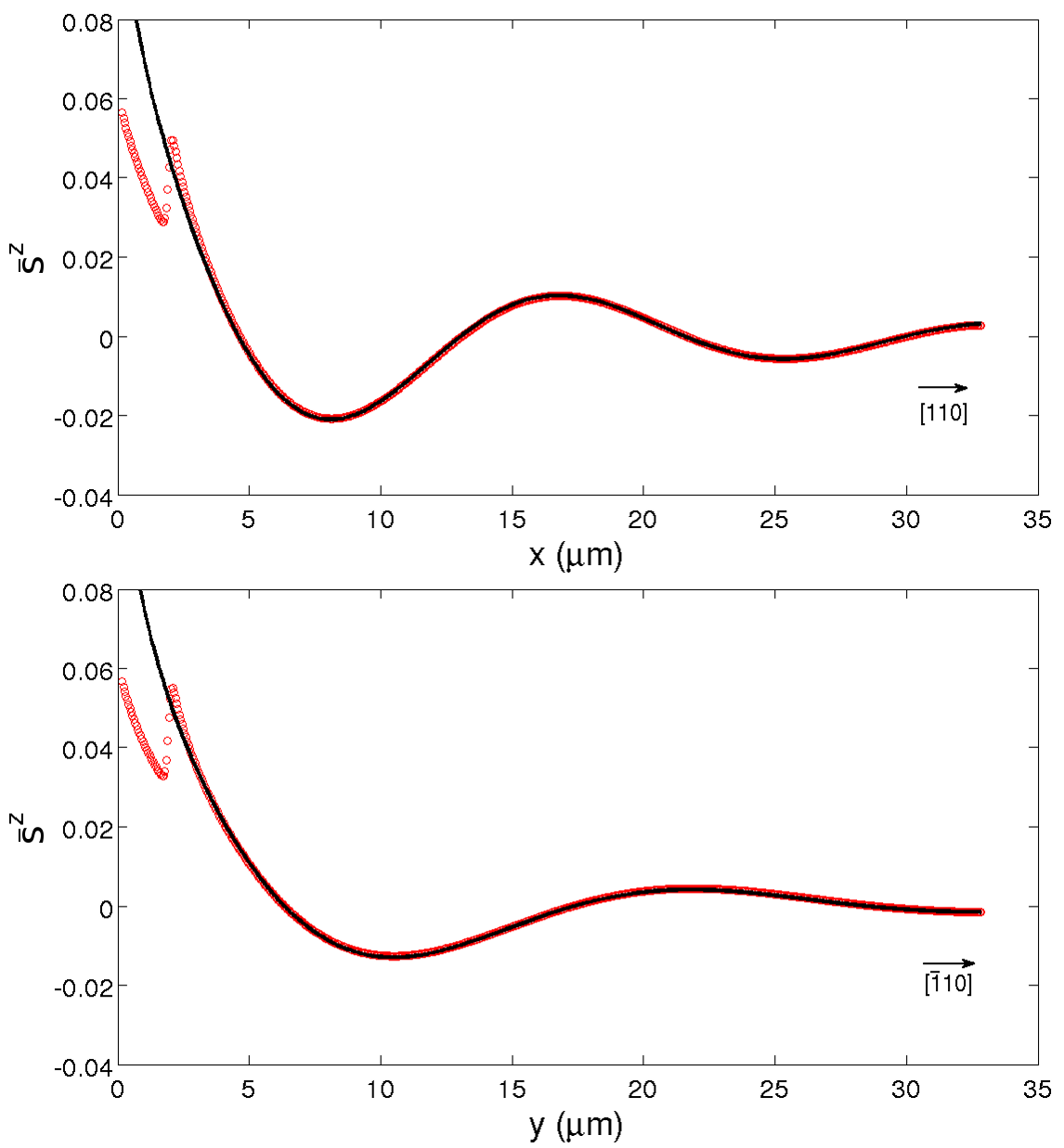}
\caption{Time-integrated spin density, $ \overline{s}^z $, along the [110] and [$\overline{1}$10] directions. 
The red circles represent the numerical solution of Eq.~\eqref{eq_cont}.  The black solid line 
shows the analytical expression \eqref{polz}.}
\label{fig_zspin}
\end{center}
\end{figure}

The time-integrated $ s^z $ takes a very simple form
along certain directions.  For example, along the $x$ direction for $y=0$ (recall that our coordinate
choice means
$\hat{\bf x}\parallel[110]$, $\hat{\bf y}\parallel[\bar{1}10]$) we find
\begin{equation}
\overline{s}^z= a_0 \frac{\exp\left(-(x-x_0)/L_{s,110}\right)}{\sqrt{y_0^2+2D(x-x_0)/v}}
\cos\left[\frac{2\pi (x-x_0)}{\lambda_{110}}\right],
\label{polz}
\end{equation}
where
\begin{equation}
\label{L110}
L_{s,110}= {v}/{2Dm^2\am^2},\; \lambda_{110}= {v}/{2\sqrt{v^2m^2\ap^2-D^2m^4\am^4}};
\end{equation}
this is plotted in Fig.~\ref{fig_zspin}, upper panel (solid black line).
The constant $a_0$ is fixed by fitting the numerical data, as discussed below.
For a SAW propagating in $y$ direction (for $ x=0 $), one finds a similar expression, with the substitutions 
$x, L_{s,110}, \lambda_{110} \rightarrow y, L_{s,\bar{1}10}, \lambda_{\bar{1}10}$:
\begin{equation}
\label{Lbar110}
L_{s,\bar{1}10}= {v}/{2Dm^2\ap^2},\; \lambda_{\bar{1}10}= {v}/{2\sqrt{v^2m^2\am^2-D^2m^4\ap^4}},
\end{equation}
compare Fig.~\ref{fig_zspin}, lower panel (solid black line).
In both propagation directions the numerical and analytical data are in good agreement for
$x,\; y\gtrsim 3\, \mu \rm m $. The reason for the deviation near the origin is that for the chosen
parameters, the standard deviation  1 $ \mathrm{\mu m}$ of the initial Gaussian is only marginally
smaller than the SAW wavelength $ 2\pi/k=2.55\,\mathrm{\mu m} $, leading to two small wires instead
of one. This causes the peak for $x,\; y$ close to this value. The spin dynamics is, however, in
both wires the same. We emphasize that the dependence of the spin precession length on the direction
of motion of the SAW is in very good agreement with the experimental observations.\cite{Sanada2011} 


\section{Miscellaneous}
\label{sec_superplus}

\subsection{Other growth directions}

Our treatment is based on the general $SU(2)$-covariant equations \eqref{eq_cont} and \eqref{eq_current}.  
The latter require as only input the specific form of the spin-orbit interaction, i.e., of the non-Abelian
vector potential $\bcA$, and yield at once the spin diffusion (Bloch) equations \eqref{eq_bloch}.
Therefore any linear-in-momentum spin-orbit term can be handled straightforwardly.
Let us consider, as another example, the $[110]$-grown GaAs quantum well experimentally studied 
in Refs.~\onlinecite{Couto2007,Couto2008}. The Rashba interaction is unchanged, compare 
Eqs.~\eqref{eq_hR} and \eqref{eq_AR}, whereas the Dresselhaus term points out-of-plane,\cite{Sih2005}
\begin{equation}
H_{so}^D = \beta p_y \sigma^z,
\end{equation}
i.e., the only non-zero component of the vector potential $\bcA_D$ is $(\cA_D)^z_y=2m\beta$.
If only the $[110]$ Dresselhaus term were present, $s^z$ would be a conserved quantity,
\cite{Hankiewicz2006, Raimondi2009} and confinement along the $x$ direction would be inconsequential.
This changes when the Rashba interaction is also taken into account. 
The eigenvalues of the $\hat\Gamma$ matrix become \cite{Raimondi2009}
\begin{eqnarray}
\Gamma_1 &=&\, 4Dm^2 \alpha^2 \,, \\ 
\Gamma_2 &=&\, 4Dm^2 \left(\alpha^2+\beta^2\right) \,, \\
\Gamma_3 &=&\, 4Dm^2  \left(2\alpha^2+\beta^2\right)\,,
\end{eqnarray}
with two eigenmode directions depending on the relative strength of the Rashba and Dresselhaus
interactions:
\begin{eqnarray}
\hat{e}_1 & \parallel & (-\alpha,0,\beta)\,,\\
\hat{e}_2 & \parallel & (0,1,0)\,,\\
\hat{e}_3 & \parallel & (\beta,0,\alpha)\,.
\end{eqnarray}
The influence of a SAW on the spin lifetimes now crucially depends on the propagation direction. 
For an $x$-propagating SAW, in the co-moving frame and after gauging away $(\cA)^y_x$ as before,
we find the eigenvalues of the $\hat\gamma$ matrix to be given by
\begin{eqnarray}
\gamma_{1,3}& = & 2Dm^2  \left(3\alpha^2+\beta^2\right)\nonumber \\
& &\pm 2\,\mathrm{i}\sqrt{v^2 m^2 \alpha^2-D^2 m^4 \left(\alpha^2+\beta^2\right)}\,,\\ 
\gamma_2 & = & \, 4Dm^2 \left(\alpha^2+\beta^2\right)\,,
\end{eqnarray}
with the eigenmode directions
\begin{eqnarray}
\hat{e}'_1 & \parallel & (-8Dm^2\alpha^2 + \gamma_1, 0, 2m\alpha v +4Dm^2\alpha\beta)\,,\\
\hat{e}'_2 & \parallel & (0,1,0)\,, \\
\hat{e}'_3 & \parallel & (-8Dm^2\alpha^2 - \gamma_3, 0, 2m\alpha v +4Dm^2\alpha\beta)\,.
\end{eqnarray}
The $y$-polarized spin eigenmode keeps its direction, $\hat{e}'_2=\hat{e}_2$, 
and its lifetime, $\gamma_2=\Gamma_2$, as in the case of a [001]-grown quantum well
(see \eqref{Gamma2} and \eqref{eig110-2}). On the other hand, the $\Gamma_1$- and $\Gamma_3$-modes
are mixed by the SAW-induced dynamics.  
By comparing $ \Gamma_{1,3} $ with the real parts of $ \gamma_{1,3} $, one sees that 
${\Re (\gamma_1)} > \Gamma_1$, i.e., the new $\gamma_1$ eigenmode has actually a shorter lifetime compared to the old one.
On the other hand, ${\Re (\gamma_3]} < \Gamma_3$, with the eigenmode lifetime increasing by a factor of two 
for strong Dresselhaus interaction, $\beta\gg\alpha$.

Even more interestingly, for a $y$-propagating SAW the relaxation is independent of $\beta$. 
The eigenvalues of the $ \hat\gamma $ matrix, in the moving frame and after the usual gauge transformation,
read
\begin{eqnarray}
\gamma_1 &=&\, 4Dm^2 \alpha^2\,,\\ 
\gamma_{2,3} &=&\, 2Dm^2  \alpha^2\nonumber \\
& &\pm 2\,\mathrm{i}\sqrt{v^2 m^2  \left(\alpha^2+\beta^2\right)-D^2 m^4 \alpha^4}\,,
\end{eqnarray}
whereas the eigenmode directions are
\begin{eqnarray}
\hat{e}'_1 & \parallel & (-\alpha,0,\beta)\,,\\
\hat{e}'_2 & \parallel & (2mv\beta, \gamma_2 , 2mv \alpha)\,,\\
\hat{e}'_3 & \parallel & (2mv\beta, -\gamma_3 , 2mv \alpha)\,.
\end{eqnarray}
Now the $\Gamma_1$-mode keeps both its lifetime, $\gamma_1=\Gamma_1$,
and its direction, $\hat{e}'_1=\hat{e}_1$, while the other two modes are strongly influenced
by the presence of the SAW.
In particular, compared to the eigenmodes $\Gamma_{2,3}$, the new eigenmodes $\gamma_{2,3}$ 
have a spin lifetime enhanced by a factor of
$2\beta^2/\alpha^2$ if the Dresselhaus interaction dominates, $\beta\gg\alpha$.

\subsection{Additional spin-orbit interactions}

Additional sources of Rashba or Dresselhaus-like spin-orbit terms are the out-of-plane 
(i.e., parallel to the quantum well growth direction) SAW field and strain.  
Experimental observations suggest these dynamical contributions
to be subleading compared to the static ones, though not completely negligible, 
especially for very strong SAW power.\cite{Sanada2011}
For this discussion we consider the non-rotated coordinates, cf. Sec.~\ref{sec_model}.

In the laboratory reference frame the additional spin-orbit interactions appear as time- and space-dependent
Rashba or Dresselhaus terms.  For example, considering a SAW propagating along the $x$ direction, Eq.~\eqref{eq_hR} is modified to 
\begin{equation}
\label{eq_hR_new}
H^R_{\rm so} = \left[\alpha+\alpha_{\rm piezo}(x,t)+\alpha_{\rm strain}(x,t)\right]
 (p_y\sigma^x-p_x\sigma^y),
\end{equation}
and similarly for the Dresselhaus terms.  In the color language this means
that we deal with a space- and time-dependent vector potential $\bcA(x,t)$. Nevertheless,
as long as the spatial variations of the spin-orbit fields are slow on the scale of the
Fermi wavelength, the $SU(2)$ approach can be employed directly, as it treats homogeneous/static
spin-orbit terms on the same footing as inhomogeneous/time-dependent ones.\cite{Gorini2010}
The dynamical nature of these additional spin-orbit interactions substantially complicates
the problem, but once more a change to the SAW co-moving reference frame offers a great simplification:  
when all disturbances, i.e., in- or out-of-plane fields, either piezoelectric or
due to strain, propagate approximately with the same sound velocity $v$,
all their contributions become static in the SAW co-moving frame,
$\bcA(x,t)\rightarrow\bcA(x)$.  Moreover, in the surfing regime 
when the carriers are confined, the vector potential can be approximated by its value at 
$x_0=\arccos\left(v/\mu E\right)/k$ (see Sec.~\ref{sec_charge}),
$\bcA(x)\approx\bcA(x_0)$.  Hence we are back to the situation discussed
in Sec.~\ref{sec_spin}, with the following modifications:
\begin{eqnarray}
\bcA_R &\rightarrow& \bcA_R+\bcA^{\rm piezo}_R(x_0)+\bcA^{\rm strain}_R(x_0)
\\
\bcA_D &\rightarrow& \bcA_D+\bcA^{\rm strain}_D(x_0).
\end{eqnarray}
This corroborates and fully justifies the intuition behind the estimations of 
$\alpha_{\rm piezo}$, $\alpha_{\rm strain}$, and $\beta_{\rm piezo}$ described in Ref.~\onlinecite{Sanada2011}.

Finally, we briefly discuss extrinsic spin relaxation, i.e., due to spin-orbit interaction
with the disorder potential $V({\bf r})$.  Extrinsic mechanisms can be included
in the color approach,\cite{Raimondi2012} and in the present case they lead to an additional
(diagonal) term $\hat\Gamma_{\rm extr}$ in the relaxation matrix $\hat\Gamma$ of Eq.~\eqref{eq_bloch},
\begin{equation}
\hat\Gamma_{\rm extr}=\frac{1}{\tau_{EY}} \, {\rm diag}(1,1,0) \, .
\end{equation} 
The Elliot-Yafet spin-flip rate $1/\tau_{EY}$ typically is negligible compared to the Dyakonov-Perel rate
(see Ref.~\onlinecite{Raimondi2009} for details), and independent of the presence of SAWs
or of confinement.  Nevertheless, a discussion focused on its role in a moving quantum dot in the presence
of a Zeeman field is given in Ref.~\onlinecite{Huang2013}.  
Note that in case the impurity potential $V(\br)$
fluctuates also out-of-plane,\cite{Dugaev2010} an Elliot-Yafet relaxation rate
for the $z$ spin component will appear.

\section{Conclusion}
\label{sec_conclusions}

By utilizing the microscopic model of a disordered two dimensional electron gas, 
we studied the effects of surface acoustic wave on the charge and spin dynamics of
photo-excited carriers, focusing on intrinsic spin-orbit mechanisms (Dyakonov-Perel relaxation). 
A SAW has to be strong enough ($ \mu E>v $) in order to transport the carriers
at the speed of sound $v$ across the sample. In this surfing regime,
the spin lifetime is considerably increased due to motional narrowing, up to a factor of two
in (001) quantum wells. The dynamics can be most conveniently described in a reference
frame co-moving with the SAW. 
In particular, we determined the SAW-induced modifications of the spin relaxation and
precession lengths. Considering also diffusion along the SAW wave front, we obtained
very good agreement with recent experimental observations.\cite{Sanada2011}  
Additional dynamical sources of spin-orbit relaxation (out-of-plane SAW field, strain)
were also shown to be most conveniently handled in the SAW co-moving frame. These effects are expected to be relevant for the ``moving quantum dots'' produced by the interference of two orthogonal SAW beams. \cite{Stotz2005,Sanada2011}

\acknowledgments{We acknowledge useful discussion with H.~Krenner and A.~Wixforth, 
as well as financial support from the German Research Foundation (DFG) through TRR 80,
and from the CEA through the DSM-Energy Program (project E112-7-Meso-Therm-DSM).}

\bibliography{Biblio-new}

\begin{thebibliography}{36}%
\makeatletter
\providecommand \@ifxundefined [1]{%
 \@ifx{#1\undefined}
}%
\providecommand \@ifnum [1]{%
 \ifnum #1\expandafter \@firstoftwo
 \else \expandafter \@secondoftwo
 \fi
}%
\providecommand \@ifx [1]{%
 \ifx #1\expandafter \@firstoftwo
 \else \expandafter \@secondoftwo
 \fi
}%
\providecommand \natexlab [1]{#1}%
\providecommand \enquote  [1]{``#1''}%
\providecommand \bibnamefont  [1]{#1}%
\providecommand \bibfnamefont [1]{#1}%
\providecommand \citenamefont [1]{#1}%
\providecommand \href@noop [0]{\@secondoftwo}%
\providecommand \href [0]{\begingroup \@sanitize@url \@href}%
\providecommand \@href[1]{\@@startlink{#1}\@@href}%
\providecommand \@@href[1]{\endgroup#1\@@endlink}%
\providecommand \@sanitize@url [0]{\catcode `\\12\catcode `\$12\catcode
  `\&12\catcode `\#12\catcode `\^12\catcode `\_12\catcode `\%12\relax}%
\providecommand \@@startlink[1]{}%
\providecommand \@@endlink[0]{}%
\providecommand \url  [0]{\begingroup\@sanitize@url \@url }%
\providecommand \@url [1]{\endgroup\@href {#1}{\urlprefix }}%
\providecommand \urlprefix  [0]{URL }%
\providecommand \Eprint [0]{\href }%
\providecommand \doibase [0]{http://dx.doi.org/}%
\providecommand \selectlanguage [0]{\@gobble}%
\providecommand \bibinfo  [0]{\@secondoftwo}%
\providecommand \bibfield  [0]{\@secondoftwo}%
\providecommand \translation [1]{[#1]}%
\providecommand \BibitemOpen [0]{}%
\providecommand \bibitemStop [0]{}%
\providecommand \bibitemNoStop [0]{.\EOS\space}%
\providecommand \EOS [0]{\spacefactor3000\relax}%
\providecommand \BibitemShut  [1]{\csname bibitem#1\endcsname}%
\let\auto@bib@innerbib\@empty
\bibitem [{\citenamefont {Stotz}\ \emph {et~al.}(2005)\citenamefont {Stotz},
  \citenamefont {Hey}, \citenamefont {Santos},\ and\ \citenamefont
  {Ploog}}]{Stotz2005}%
  \BibitemOpen
  \bibfield  {author} {\bibinfo {author} {\bibfnamefont {J.~A.~H.}\
  \bibnamefont {Stotz}}, \bibinfo {author} {\bibfnamefont {R.}~\bibnamefont
  {Hey}}, \bibinfo {author} {\bibfnamefont {P.~V.}\ \bibnamefont {Santos}}, \
  and\ \bibinfo {author} {\bibfnamefont {K.~H.}\ \bibnamefont {Ploog}},\
  }\href@noop {} {\bibfield  {journal} {\bibinfo  {journal} {Nature Mat.}\
  }\textbf {\bibinfo {volume} {4}},\ \bibinfo {pages} {585} (\bibinfo {year}
  {2005})}\BibitemShut {NoStop}%
\bibitem [{\citenamefont {Awschalom}\ and\ \citenamefont
  {Flatt{\'e}}(2007)}]{Awschalom2007}%
  \BibitemOpen
  \bibfield  {author} {\bibinfo {author} {\bibfnamefont {D.~D.}\ \bibnamefont
  {Awschalom}}\ and\ \bibinfo {author} {\bibfnamefont {M.~E.}\ \bibnamefont
  {Flatt{\'e}}},\ }\href@noop {} {\bibfield  {journal} {\bibinfo  {journal}
  {Nature Phys.}\ }\textbf {\bibinfo {volume} {3}},\ \bibinfo {pages} {153}
  (\bibinfo {year} {2007})}\BibitemShut {NoStop}%
\bibitem [{\citenamefont {Sogawa}\ \emph {et~al.}(2001)\citenamefont {Sogawa},
  \citenamefont {Santos}, \citenamefont {Zhang}, \citenamefont {Eshlaghi},
  \citenamefont {Wieck},\ and\ \citenamefont {Ploog}}]{Sogawa2001}%
  \BibitemOpen
  \bibfield  {author} {\bibinfo {author} {\bibfnamefont {T.}~\bibnamefont
  {Sogawa}}, \bibinfo {author} {\bibfnamefont {P.~V.}\ \bibnamefont {Santos}},
  \bibinfo {author} {\bibfnamefont {S.-C.}\ \bibnamefont {Zhang}}, \bibinfo
  {author} {\bibfnamefont {S.}~\bibnamefont {Eshlaghi}}, \bibinfo {author}
  {\bibfnamefont {A.~D.}\ \bibnamefont {Wieck}}, \ and\ \bibinfo {author}
  {\bibfnamefont {K.~H.}\ \bibnamefont {Ploog}},\ }\href@noop {} {\bibfield
  {journal} {\bibinfo  {journal} {Phys. Rev. Lett.}\ }\textbf {\bibinfo
  {volume} {87}},\ \bibinfo {pages} {276601} (\bibinfo {year}
  {2001})}\BibitemShut {NoStop}%
\bibitem [{\citenamefont {Couto}\ \emph {et~al.}(2007)\citenamefont {Couto},
  \citenamefont {Iikawa}, \citenamefont {Rudolph}, \citenamefont {Hey},\ and\
  \citenamefont {Santos}}]{Couto2007}%
  \BibitemOpen
  \bibfield  {author} {\bibinfo {author} {\bibfnamefont {O.}~\bibnamefont
  {Couto}}, \bibinfo {author} {\bibfnamefont {F.}~\bibnamefont {Iikawa}},
  \bibinfo {author} {\bibfnamefont {J.}~\bibnamefont {Rudolph}}, \bibinfo
  {author} {\bibfnamefont {R.}~\bibnamefont {Hey}}, \ and\ \bibinfo {author}
  {\bibfnamefont {P.~V.}\ \bibnamefont {Santos}},\ }\href@noop {} {\bibfield
  {journal} {\bibinfo  {journal} {Phys. Rev. Lett.}\ }\textbf {\bibinfo
  {volume} {98}},\ \bibinfo {pages} {036603} (\bibinfo {year}
  {2007})}\BibitemShut {NoStop}%
\bibitem [{\citenamefont {Couto}\ \emph {et~al.}(2008)\citenamefont {Couto},
  \citenamefont {Hey},\ and\ \citenamefont {Santos}}]{Couto2008}%
  \BibitemOpen
  \bibfield  {author} {\bibinfo {author} {\bibfnamefont {O.}~\bibnamefont
  {Couto}}, \bibinfo {author} {\bibfnamefont {R.}~\bibnamefont {Hey}}, \ and\
  \bibinfo {author} {\bibfnamefont {P.~V.}\ \bibnamefont {Santos}},\
  }\href@noop {} {\bibfield  {journal} {\bibinfo  {journal} {Phys. Rev. B}\
  }\textbf {\bibinfo {volume} {78}},\ \bibinfo {pages} {153305} (\bibinfo
  {year} {2008})}\BibitemShut {NoStop}%
\bibitem [{\citenamefont {Sanada}\ \emph {et~al.}(2011)\citenamefont {Sanada},
  \citenamefont {Sogawa}, \citenamefont {Gotoh}, \citenamefont {Onomitsu},
  \citenamefont {Kohda}, \citenamefont {Nitta},\ and\ \citenamefont
  {Santos}}]{Sanada2011}%
  \BibitemOpen
  \bibfield  {author} {\bibinfo {author} {\bibfnamefont {H.}~\bibnamefont
  {Sanada}}, \bibinfo {author} {\bibfnamefont {T.}~\bibnamefont {Sogawa}},
  \bibinfo {author} {\bibfnamefont {H.}~\bibnamefont {Gotoh}}, \bibinfo
  {author} {\bibfnamefont {K.}~\bibnamefont {Onomitsu}}, \bibinfo {author}
  {\bibfnamefont {M.}~\bibnamefont {Kohda}}, \bibinfo {author} {\bibfnamefont
  {J.}~\bibnamefont {Nitta}}, \ and\ \bibinfo {author} {\bibfnamefont {P.~V.}\
  \bibnamefont {Santos}},\ }\href@noop {} {\bibfield  {journal} {\bibinfo
  {journal} {Phys. Rev. Lett.}\ }\textbf {\bibinfo {volume} {106}},\ \bibinfo
  {pages} {216602} (\bibinfo {year} {2011})}\BibitemShut {NoStop}%
\bibitem [{\citenamefont {Bir}\ \emph {et~al.}(1975)\citenamefont {Bir},
  \citenamefont {Aronov},\ and\ \citenamefont {Pikus}}]{Bir1975}%
  \BibitemOpen
  \bibfield  {author} {\bibinfo {author} {\bibfnamefont {G.~L.}\ \bibnamefont
  {Bir}}, \bibinfo {author} {\bibfnamefont {A.~G.}\ \bibnamefont {Aronov}}, \
  and\ \bibinfo {author} {\bibfnamefont {G.~E.}\ \bibnamefont {Pikus}},\
  }\href@noop {} {\bibfield  {journal} {\bibinfo  {journal} {Sov. Phys. JETP}\
  }\textbf {\bibinfo {volume} {42}},\ \bibinfo {pages} {705} (\bibinfo {year}
  {1975})}\BibitemShut {NoStop}%
\bibitem [{\citenamefont {Dyakonov}\ and\ \citenamefont
  {Perel}(1972)}]{Dyakonov1972}%
  \BibitemOpen
  \bibfield  {author} {\bibinfo {author} {\bibfnamefont {M.}~\bibnamefont
  {Dyakonov}}\ and\ \bibinfo {author} {\bibfnamefont {V.}~\bibnamefont
  {Perel}},\ }\href@noop {} {\bibfield  {journal} {\bibinfo  {journal} {Sov.
  Phys. Solid State}\ }\textbf {\bibinfo {volume} {13}},\ \bibinfo {pages}
  {3023} (\bibinfo {year} {1972})}\BibitemShut {NoStop}%
\bibitem [{\citenamefont {Holleitner}\ \emph {et~al.}(2006)\citenamefont
  {Holleitner}, \citenamefont {Sih}, \citenamefont {Myers}, \citenamefont
  {Gossard},\ and\ \citenamefont {Awschalom}}]{Holleitner2006}%
  \BibitemOpen
  \bibfield  {author} {\bibinfo {author} {\bibfnamefont {A.~W.}\ \bibnamefont
  {Holleitner}}, \bibinfo {author} {\bibfnamefont {V.}~\bibnamefont {Sih}},
  \bibinfo {author} {\bibfnamefont {R.~C.}\ \bibnamefont {Myers}}, \bibinfo
  {author} {\bibfnamefont {A.~C.}\ \bibnamefont {Gossard}}, \ and\ \bibinfo
  {author} {\bibfnamefont {D.~D.}\ \bibnamefont {Awschalom}},\ }\href@noop {}
  {\bibfield  {journal} {\bibinfo  {journal} {Phys. Rev. Lett.}\ }\textbf
  {\bibinfo {volume} {97}},\ \bibinfo {pages} {036805} (\bibinfo {year}
  {2006})}\BibitemShut {NoStop}%
\bibitem [{\citenamefont {Schwab}\ \emph {et~al.}(2006)\citenamefont {Schwab},
  \citenamefont {Dzierzawa}, \citenamefont {Gorini},\ and\ \citenamefont
  {Raimondi}}]{Schwab2006}%
  \BibitemOpen
  \bibfield  {author} {\bibinfo {author} {\bibfnamefont {P.}~\bibnamefont
  {Schwab}}, \bibinfo {author} {\bibfnamefont {M.}~\bibnamefont {Dzierzawa}},
  \bibinfo {author} {\bibfnamefont {C.}~\bibnamefont {Gorini}}, \ and\ \bibinfo
  {author} {\bibfnamefont {R.}~\bibnamefont {Raimondi}},\ }\href@noop {}
  {\bibfield  {journal} {\bibinfo  {journal} {Phys. Rev. B}\ }\textbf {\bibinfo
  {volume} {74}},\ \bibinfo {pages} {155316} (\bibinfo {year}
  {2006})}\BibitemShut {NoStop}%
\bibitem [{\citenamefont {Raimondi}\ and\ \citenamefont
  {Schwab}(2009)}]{Raimondi2009}%
  \BibitemOpen
  \bibfield  {author} {\bibinfo {author} {\bibfnamefont {R.}~\bibnamefont
  {Raimondi}}\ and\ \bibinfo {author} {\bibfnamefont {P.}~\bibnamefont
  {Schwab}},\ }\href@noop {} {\bibfield  {journal} {\bibinfo  {journal} {EPL}\
  }\textbf {\bibinfo {volume} {87}},\ \bibinfo {pages} {37008} (\bibinfo {year}
  {2009})}\BibitemShut {NoStop}%
\bibitem [{\citenamefont {Merkulov}\ \emph {et~al.}(2002)\citenamefont
  {Merkulov}, \citenamefont {Efros},\ and\ \citenamefont
  {Rosen}}]{Merkulov2002}%
  \BibitemOpen
  \bibfield  {author} {\bibinfo {author} {\bibfnamefont {I.~A.}\ \bibnamefont
  {Merkulov}}, \bibinfo {author} {\bibfnamefont {A.~L.}\ \bibnamefont {Efros}},
  \ and\ \bibinfo {author} {\bibfnamefont {M.}~\bibnamefont {Rosen}},\
  }\href@noop {} {\bibfield  {journal} {\bibinfo  {journal} {Phys. Rev. B}\
  }\textbf {\bibinfo {volume} {65}},\ \bibinfo {pages} {205309} (\bibinfo
  {year} {2002})}\BibitemShut {NoStop}%
\bibitem [{\citenamefont {Braun}\ \emph {et~al.}(2005)\citenamefont {Braun},
  \citenamefont {Marie}, \citenamefont {Lombez}, \citenamefont {Urbaszek},
  \citenamefont {Amand}, \citenamefont {Renucci}, \citenamefont {Kalevich},
  \citenamefont {Kavokin}, \citenamefont {Krebs}, \citenamefont {Voisin},\ and\
  \citenamefont {Masumoto}}]{Braun2005}%
  \BibitemOpen
  \bibfield  {author} {\bibinfo {author} {\bibfnamefont {P.-F.}\ \bibnamefont
  {Braun}}, \bibinfo {author} {\bibfnamefont {X.}~\bibnamefont {Marie}},
  \bibinfo {author} {\bibfnamefont {L.}~\bibnamefont {Lombez}}, \bibinfo
  {author} {\bibfnamefont {B.}~\bibnamefont {Urbaszek}}, \bibinfo {author}
  {\bibfnamefont {T.}~\bibnamefont {Amand}}, \bibinfo {author} {\bibfnamefont
  {P.}~\bibnamefont {Renucci}}, \bibinfo {author} {\bibfnamefont {V.~K.}\
  \bibnamefont {Kalevich}}, \bibinfo {author} {\bibfnamefont {K.~V.}\
  \bibnamefont {Kavokin}}, \bibinfo {author} {\bibfnamefont {O.}~\bibnamefont
  {Krebs}}, \bibinfo {author} {\bibfnamefont {P.}~\bibnamefont {Voisin}}, \
  and\ \bibinfo {author} {\bibfnamefont {Y.}~\bibnamefont {Masumoto}},\
  }\href@noop {} {\bibfield  {journal} {\bibinfo  {journal} {Phys. Rev. Lett.}\
  }\textbf {\bibinfo {volume} {94}},\ \bibinfo {pages} {116601} (\bibinfo
  {year} {2005})}\BibitemShut {NoStop}%
\bibitem [{\citenamefont {Echeverr{\'i}a-Arrondo}\ and\ \citenamefont
  {Sherman}(2013)}]{Echeverria2013}%
  \BibitemOpen
  \bibfield  {author} {\bibinfo {author} {\bibfnamefont {C.}~\bibnamefont
  {Echeverr{\'i}a-Arrondo}}\ and\ \bibinfo {author} {\bibfnamefont {E.~Y.}\
  \bibnamefont {Sherman}},\ }\href@noop {} {\bibfield  {journal} {\bibinfo
  {journal} {Phys. Rev. B}\ }\textbf {\bibinfo {volume} {87}},\ \bibinfo
  {pages} {081410(R)} (\bibinfo {year} {2013})}\BibitemShut {NoStop}%
\bibitem [{\citenamefont {Rashba}\ and\ \citenamefont
  {Bychkov}(1984)}]{Bychkov1984}%
  \BibitemOpen
  \bibfield  {author} {\bibinfo {author} {\bibfnamefont {E.~I.}\ \bibnamefont
  {Rashba}}\ and\ \bibinfo {author} {\bibfnamefont {Y.}~\bibnamefont
  {Bychkov}},\ }\href@noop {} {\bibfield  {journal} {\bibinfo  {journal} {J.
  Phys. C}\ }\textbf {\bibinfo {volume} {17}},\ \bibinfo {pages} {6039}
  (\bibinfo {year} {1984})}\BibitemShut {NoStop}%
\bibitem [{\citenamefont {Dresselhaus}(1955)}]{Dresselhaus1955}%
  \BibitemOpen
  \bibfield  {author} {\bibinfo {author} {\bibfnamefont {G.}~\bibnamefont
  {Dresselhaus}},\ }\href@noop {} {\bibfield  {journal} {\bibinfo  {journal}
  {Phys. Rev.}\ }\textbf {\bibinfo {volume} {100}},\ \bibinfo {pages} {580}
  (\bibinfo {year} {1955})}\BibitemShut {NoStop}%
\bibitem [{\citenamefont {Bernevig}\ and\ \citenamefont
  {Zhang}(2005)}]{Bernevig2005}%
  \BibitemOpen
  \bibfield  {author} {\bibinfo {author} {\bibfnamefont {B.}~\bibnamefont
  {Bernevig}}\ and\ \bibinfo {author} {\bibfnamefont {S.-C.}\ \bibnamefont
  {Zhang}},\ }\href@noop {} {\bibfield  {journal} {\bibinfo  {journal} {Phys.
  Rev. B}\ }\textbf {\bibinfo {volume} {72}},\ \bibinfo {pages} {115204}
  (\bibinfo {year} {2005})}\BibitemShut {NoStop}%
\bibitem [{\citenamefont {Studer}\ \emph {et~al.}(2010)\citenamefont {Studer},
  \citenamefont {Walser}, \citenamefont {Baer}, \citenamefont {Rusterholz},
  \citenamefont {Sch{\"o}n}, \citenamefont {Schuh}, \citenamefont
  {Wegscheider}, \citenamefont {Ensslin},\ and\ \citenamefont
  {Salis}}]{Studer2010}%
  \BibitemOpen
  \bibfield  {author} {\bibinfo {author} {\bibfnamefont {M.}~\bibnamefont
  {Studer}}, \bibinfo {author} {\bibfnamefont {M.~P.}\ \bibnamefont {Walser}},
  \bibinfo {author} {\bibfnamefont {S.}~\bibnamefont {Baer}}, \bibinfo {author}
  {\bibfnamefont {H.}~\bibnamefont {Rusterholz}}, \bibinfo {author}
  {\bibfnamefont {S.}~\bibnamefont {Sch{\"o}n}}, \bibinfo {author}
  {\bibfnamefont {D.}~\bibnamefont {Schuh}}, \bibinfo {author} {\bibfnamefont
  {W.}~\bibnamefont {Wegscheider}}, \bibinfo {author} {\bibfnamefont
  {K.}~\bibnamefont {Ensslin}}, \ and\ \bibinfo {author} {\bibfnamefont
  {G.}~\bibnamefont {Salis}},\ }\href@noop {} {\bibfield  {journal} {\bibinfo
  {journal} {Phys. Rev. B}\ }\textbf {\bibinfo {volume} {82}},\ \bibinfo
  {pages} {235320} (\bibinfo {year} {2010})}\BibitemShut {NoStop}%
\bibitem [{\citenamefont {Walser}\ \emph {et~al.}(2012)\citenamefont {Walser},
  \citenamefont {Siegenthaler}, \citenamefont {Lechner}, \citenamefont {Schuh},
  \citenamefont {Ganichev}, \citenamefont {Wegscheider},\ and\ \citenamefont
  {Salis}}]{Walser2012}%
  \BibitemOpen
  \bibfield  {author} {\bibinfo {author} {\bibfnamefont {M.~P.}\ \bibnamefont
  {Walser}}, \bibinfo {author} {\bibfnamefont {U.}~\bibnamefont
  {Siegenthaler}}, \bibinfo {author} {\bibfnamefont {V.}~\bibnamefont
  {Lechner}}, \bibinfo {author} {\bibfnamefont {D.}~\bibnamefont {Schuh}},
  \bibinfo {author} {\bibfnamefont {S.~D.}\ \bibnamefont {Ganichev}}, \bibinfo
  {author} {\bibfnamefont {W.}~\bibnamefont {Wegscheider}}, \ and\ \bibinfo
  {author} {\bibfnamefont {G.}~\bibnamefont {Salis}},\ }\href@noop {}
  {\bibfield  {journal} {\bibinfo  {journal} {Phys. Rev. B}\ }\textbf {\bibinfo
  {volume} {86}},\ \bibinfo {pages} {195309} (\bibinfo {year}
  {2012})}\BibitemShut {NoStop}%
\bibitem [{\citenamefont {Mathur}\ and\ \citenamefont
  {Stone}(1992)}]{Mathur1992}%
  \BibitemOpen
  \bibfield  {author} {\bibinfo {author} {\bibfnamefont {H.}~\bibnamefont
  {Mathur}}\ and\ \bibinfo {author} {\bibfnamefont {A.~D.}\ \bibnamefont
  {Stone}},\ }\href@noop {} {\bibfield  {journal} {\bibinfo  {journal} {Phys.
  Rev. Lett.}\ }\textbf {\bibinfo {volume} {68}},\ \bibinfo {pages} {2964}
  (\bibinfo {year} {1992})}\BibitemShut {NoStop}%
\bibitem [{\citenamefont {Fr{\"o}hlich}\ and\ \citenamefont
  {Studer}(1993)}]{Frohlich1993}%
  \BibitemOpen
  \bibfield  {author} {\bibinfo {author} {\bibfnamefont {J.}~\bibnamefont
  {Fr{\"o}hlich}}\ and\ \bibinfo {author} {\bibfnamefont {U.~M.}\ \bibnamefont
  {Studer}},\ }\href@noop {} {\bibfield  {journal} {\bibinfo  {journal} {Rev.
  Mod. Phys.}\ }\textbf {\bibinfo {volume} {65}},\ \bibinfo {pages} {733}
  (\bibinfo {year} {1993})}\BibitemShut {NoStop}%
\bibitem [{\citenamefont {Tokatly}(2008)}]{Tokatly2008}%
  \BibitemOpen
  \bibfield  {author} {\bibinfo {author} {\bibfnamefont {I.~V.}\ \bibnamefont
  {Tokatly}},\ }\href@noop {} {\bibfield  {journal} {\bibinfo  {journal} {Phys.
  Rev. Lett.}\ }\textbf {\bibinfo {volume} {101}},\ \bibinfo {pages} {106601}
  (\bibinfo {year} {2008})}\BibitemShut {NoStop}%
\bibitem [{\citenamefont {Gorini}\ \emph {et~al.}(2010)\citenamefont {Gorini},
  \citenamefont {Schwab}, \citenamefont {Raimondi},\ and\ \citenamefont
  {Shelankov}}]{Gorini2010}%
  \BibitemOpen
  \bibfield  {author} {\bibinfo {author} {\bibfnamefont {C.}~\bibnamefont
  {Gorini}}, \bibinfo {author} {\bibfnamefont {P.}~\bibnamefont {Schwab}},
  \bibinfo {author} {\bibfnamefont {R.}~\bibnamefont {Raimondi}}, \ and\
  \bibinfo {author} {\bibfnamefont {A.~L.}\ \bibnamefont {Shelankov}},\
  }\href@noop {} {\bibfield  {journal} {\bibinfo  {journal} {Phys. Rev. B}\
  }\textbf {\bibinfo {volume} {82}},\ \bibinfo {pages} {195316} (\bibinfo
  {year} {2010})}\BibitemShut {NoStop}%
\bibitem [{\citenamefont {Raimondi}\ \emph {et~al.}(2012)\citenamefont
  {Raimondi}, \citenamefont {Schwab}, \citenamefont {Gorini},\ and\
  \citenamefont {Vignale}}]{Raimondi2012}%
  \BibitemOpen
  \bibfield  {author} {\bibinfo {author} {\bibfnamefont {R.}~\bibnamefont
  {Raimondi}}, \bibinfo {author} {\bibfnamefont {P.}~\bibnamefont {Schwab}},
  \bibinfo {author} {\bibfnamefont {C.}~\bibnamefont {Gorini}}, \ and\ \bibinfo
  {author} {\bibfnamefont {G.}~\bibnamefont {Vignale}},\ }\href@noop {}
  {\bibfield  {journal} {\bibinfo  {journal} {Ann. Phys. (Berlin)}\ }\textbf
  {\bibinfo {volume} {524}},\ \bibinfo {pages} {153} (\bibinfo {year}
  {2012})}\BibitemShut {NoStop}%
\bibitem [{\citenamefont {Mamishev}\ \emph {et~al.}(2004)\citenamefont
  {Mamishev}, \citenamefont {Sundara-Rajan},\ and\ \citenamefont
  {Zahn}}]{Mamishev2004}%
  \BibitemOpen
  \bibfield  {author} {\bibinfo {author} {\bibfnamefont {A.}~\bibnamefont
  {Mamishev}}, \bibinfo {author} {\bibfnamefont {K.}~\bibnamefont
  {Sundara-Rajan}}, \ and\ \bibinfo {author} {\bibfnamefont {M.}~\bibnamefont
  {Zahn}},\ }\href@noop {} {\bibfield  {journal} {\bibinfo  {journal} {Proc.
  IEEE}\ }\textbf {\bibinfo {volume} {92}},\ \bibinfo {pages} {808} (\bibinfo
  {year} {2004})}\BibitemShut {NoStop}%
\bibitem [{\citenamefont {Morgan}(2007)}]{Morgan2007}%
  \BibitemOpen
  \bibfield  {author} {\bibinfo {author} {\bibfnamefont {D.}~\bibnamefont
  {Morgan}},\ }\href@noop {} {\emph {\bibinfo {title} {Surface Acoustic Wave
  Filters}}}\ (\bibinfo  {publisher} {Academic Press},\ \bibinfo {year}
  {2007})\BibitemShut {NoStop}%
\bibitem [{\citenamefont {Cameron}\ \emph {et~al.}(1996)\citenamefont
  {Cameron}, \citenamefont {Riblet},\ and\ \citenamefont
  {Miller}}]{Cameron1996}%
  \BibitemOpen
  \bibfield  {author} {\bibinfo {author} {\bibfnamefont {A.}~\bibnamefont
  {Cameron}}, \bibinfo {author} {\bibfnamefont {P.}~\bibnamefont {Riblet}}, \
  and\ \bibinfo {author} {\bibfnamefont {A.}~\bibnamefont {Miller}},\
  }\href@noop {} {\bibfield  {journal} {\bibinfo  {journal} {Phys. Rev. Lett.}\
  }\textbf {\bibinfo {volume} {76}},\ \bibinfo {pages} {4793} (\bibinfo {year}
  {1996})}\BibitemShut {NoStop}%
\bibitem [{\citenamefont {Wang}\ \emph {et~al.}(2013)\citenamefont {Wang},
  \citenamefont {Liu}, \citenamefont {Balocchi}, \citenamefont {Renucci},
  \citenamefont {Zhu}, \citenamefont {Amand}, \citenamefont {Fontaine},\ and\
  \citenamefont {Marie}}]{Wang2013}%
  \BibitemOpen
  \bibfield  {author} {\bibinfo {author} {\bibfnamefont {G.}~\bibnamefont
  {Wang}}, \bibinfo {author} {\bibfnamefont {B.~L.}\ \bibnamefont {Liu}},
  \bibinfo {author} {\bibfnamefont {A.}~\bibnamefont {Balocchi}}, \bibinfo
  {author} {\bibfnamefont {P.}~\bibnamefont {Renucci}}, \bibinfo {author}
  {\bibfnamefont {C.~R.}\ \bibnamefont {Zhu}}, \bibinfo {author} {\bibfnamefont
  {T.}~\bibnamefont {Amand}}, \bibinfo {author} {\bibfnamefont
  {C.}~\bibnamefont {Fontaine}}, \ and\ \bibinfo {author} {\bibfnamefont
  {X.}~\bibnamefont {Marie}},\ }\href@noop {} {\bibfield  {journal} {\bibinfo
  {journal} {Nature Comm.}\ }\textbf {\bibinfo {volume} {4}},\ \bibinfo {pages}
  {2372} (\bibinfo {year} {2013})}\BibitemShut {NoStop}%
\bibitem [{\citenamefont {Garc{\'i}a-Crist{\'o}bal}\ \emph
  {et~al.}(2004)\citenamefont {Garc{\'i}a-Crist{\'o}bal}, \citenamefont
  {Cantarero}, \citenamefont {Alsina},\ and\ \citenamefont
  {Santos}}]{Garcia-Cristobal2004}%
  \BibitemOpen
  \bibfield  {author} {\bibinfo {author} {\bibfnamefont {A.}~\bibnamefont
  {Garc{\'i}a-Crist{\'o}bal}}, \bibinfo {author} {\bibfnamefont
  {A.}~\bibnamefont {Cantarero}}, \bibinfo {author} {\bibfnamefont
  {F.}~\bibnamefont {Alsina}}, \ and\ \bibinfo {author} {\bibfnamefont {P.~V.}\
  \bibnamefont {Santos}},\ }\href@noop {} {\bibfield  {journal} {\bibinfo
  {journal} {Phys. Rev. B}\ }\textbf {\bibinfo {volume} {69}},\ \bibinfo
  {pages} {205301} (\bibinfo {year} {2004})}\BibitemShut {NoStop}%
\bibitem [{Note1()}]{Note1}%
  \BibitemOpen
  \bibinfo {note} {Note that $ k x_0 $ lies in the range $ 0\protect \ldots \pi
  $. This is also apparent from Eq. \ref {eq_carrier}.}\BibitemShut {Stop}%
\bibitem [{\citenamefont {Koralek}\ \emph {et~al.}(2009)\citenamefont
  {Koralek}, \citenamefont {Weber}, \citenamefont {Orenstein}, \citenamefont
  {Bernevig}, \citenamefont {Zhang}, \citenamefont {Mack},\ and\ \citenamefont
  {Awschalom}}]{Koralek2009}%
  \BibitemOpen
  \bibfield  {author} {\bibinfo {author} {\bibfnamefont {J.~D.}\ \bibnamefont
  {Koralek}}, \bibinfo {author} {\bibfnamefont {C.~P.}\ \bibnamefont {Weber}},
  \bibinfo {author} {\bibfnamefont {J.}~\bibnamefont {Orenstein}}, \bibinfo
  {author} {\bibfnamefont {B.~A.}\ \bibnamefont {Bernevig}}, \bibinfo {author}
  {\bibfnamefont {S.-C.}\ \bibnamefont {Zhang}}, \bibinfo {author}
  {\bibfnamefont {S.}~\bibnamefont {Mack}}, \ and\ \bibinfo {author}
  {\bibfnamefont {D.~D.}\ \bibnamefont {Awschalom}},\ }\href@noop {} {\bibfield
   {journal} {\bibinfo  {journal} {Nature}\ }\textbf {\bibinfo {volume}
  {458}},\ \bibinfo {pages} {610} (\bibinfo {year} {2009})}\BibitemShut
  {NoStop}%
\bibitem [{\citenamefont {Kohda}\ \emph {et~al.}(2012)\citenamefont {Kohda},
  \citenamefont {Lechner}, \citenamefont {Kunihashi}, \citenamefont
  {Dollinger}, \citenamefont {Olbrich}, \citenamefont {Sch{\"o}nhuber},
  \citenamefont {Caspers}, \citenamefont {Bel'kov}, \citenamefont {Golub},
  \citenamefont {Weiss}, \citenamefont {Richter}, \citenamefont {Nitta},\ and\
  \citenamefont {Ganichev}}]{Kohda2012}%
  \BibitemOpen
  \bibfield  {author} {\bibinfo {author} {\bibfnamefont {M.}~\bibnamefont
  {Kohda}}, \bibinfo {author} {\bibfnamefont {V.}~\bibnamefont {Lechner}},
  \bibinfo {author} {\bibfnamefont {Y.}~\bibnamefont {Kunihashi}}, \bibinfo
  {author} {\bibfnamefont {T.}~\bibnamefont {Dollinger}}, \bibinfo {author}
  {\bibfnamefont {P.}~\bibnamefont {Olbrich}}, \bibinfo {author} {\bibfnamefont
  {C.}~\bibnamefont {Sch{\"o}nhuber}}, \bibinfo {author} {\bibfnamefont
  {I.}~\bibnamefont {Caspers}}, \bibinfo {author} {\bibfnamefont {V.~V.}\
  \bibnamefont {Bel'kov}}, \bibinfo {author} {\bibfnamefont {L.~E.}\
  \bibnamefont {Golub}}, \bibinfo {author} {\bibfnamefont {D.}~\bibnamefont
  {Weiss}}, \bibinfo {author} {\bibfnamefont {K.}~\bibnamefont {Richter}},
  \bibinfo {author} {\bibfnamefont {J.}~\bibnamefont {Nitta}}, \ and\ \bibinfo
  {author} {\bibfnamefont {S.~D.}\ \bibnamefont {Ganichev}},\ }\href@noop {}
  {\bibfield  {journal} {\bibinfo  {journal} {Phys. Rev. B}\ }\textbf {\bibinfo
  {volume} {86}},\ \bibinfo {pages} {081306} (\bibinfo {year}
  {2012})}\BibitemShut {NoStop}%
\bibitem [{\citenamefont {Sih}\ \emph {et~al.}(2005)\citenamefont {Sih},
  \citenamefont {Myers}, \citenamefont {Kato}, \citenamefont {Lau},
  \citenamefont {Gossard},\ and\ \citenamefont {Awschalom}}]{Sih2005}%
  \BibitemOpen
  \bibfield  {author} {\bibinfo {author} {\bibfnamefont {V.}~\bibnamefont
  {Sih}}, \bibinfo {author} {\bibfnamefont {R.~C.}\ \bibnamefont {Myers}},
  \bibinfo {author} {\bibfnamefont {Y.~K.}\ \bibnamefont {Kato}}, \bibinfo
  {author} {\bibfnamefont {W.~H.}\ \bibnamefont {Lau}}, \bibinfo {author}
  {\bibfnamefont {A.~C.}\ \bibnamefont {Gossard}}, \ and\ \bibinfo {author}
  {\bibfnamefont {D.~D.}\ \bibnamefont {Awschalom}},\ }\href@noop {} {\bibfield
   {journal} {\bibinfo  {journal} {Nat. Phys.}\ }\textbf {\bibinfo {volume}
  {1}},\ \bibinfo {pages} {31} (\bibinfo {year} {2005})}\BibitemShut {NoStop}%
\bibitem [{\citenamefont {Hankiewicz}\ \emph {et~al.}(2006)\citenamefont
  {Hankiewicz}, \citenamefont {Vignale},\ and\ \citenamefont
  {Flatt{\'e}}}]{Hankiewicz2006}%
  \BibitemOpen
  \bibfield  {author} {\bibinfo {author} {\bibfnamefont {E.~M.}\ \bibnamefont
  {Hankiewicz}}, \bibinfo {author} {\bibfnamefont {G.}~\bibnamefont {Vignale}},
  \ and\ \bibinfo {author} {\bibfnamefont {M.~E.}\ \bibnamefont {Flatt{\'e}}},\
  }\href@noop {} {\bibfield  {journal} {\bibinfo  {journal} {Phys. Rev. Lett.}\
  }\textbf {\bibinfo {volume} {97}},\ \bibinfo {pages} {266601} (\bibinfo
  {year} {2006})}\BibitemShut {NoStop}%
\bibitem [{\citenamefont {Huang}\ and\ \citenamefont {Hu}(2013)}]{Huang2013}%
  \BibitemOpen
  \bibfield  {author} {\bibinfo {author} {\bibfnamefont {P.}~\bibnamefont
  {Huang}}\ and\ \bibinfo {author} {\bibfnamefont {X.}~\bibnamefont {Hu}},\
  }\href@noop {} {\bibfield  {journal} {\bibinfo  {journal} {Phys. Rev. B}\
  }\textbf {\bibinfo {volume} {88}},\ \bibinfo {pages} {075301} (\bibinfo
  {year} {2013})}\BibitemShut {NoStop}%
\bibitem [{\citenamefont {Dugaev}\ \emph {et~al.}(2010)\citenamefont {Dugaev},
  \citenamefont {Inglot}, \citenamefont {Sherman},\ and\ \citenamefont
  {Barna{\'s}}}]{Dugaev2010}%
  \BibitemOpen
  \bibfield  {author} {\bibinfo {author} {\bibfnamefont {V.~K.}\ \bibnamefont
  {Dugaev}}, \bibinfo {author} {\bibfnamefont {M.}~\bibnamefont {Inglot}},
  \bibinfo {author} {\bibfnamefont {E.~Y.}\ \bibnamefont {Sherman}}, \ and\
  \bibinfo {author} {\bibfnamefont {J.}~\bibnamefont {Barna{\'s}}},\
  }\href@noop {} {\bibfield  {journal} {\bibinfo  {journal} {Phys. Rev. B}\
  }\textbf {\bibinfo {volume} {82}},\ \bibinfo {pages} {121310(R)} (\bibinfo
  {year} {2010})}\BibitemShut {NoStop}%
\end{thebibliography}%

\end{document}